\documentclass[iop, revtex4]{emulateapj}
\usepackage{natbib,graphics,graphicx,multirow,amsmath,cancel,color,xcolor,lscape,hyperref}
\usepackage{soul}
\usepackage[normalem]{ulem}
\usepackage{hyperref}
\hypersetup{
    colorlinks=true,
    linkcolor=blue,
    filecolor=magenta,      
    urlcolor=cyan,
}
 
\shorttitle{Coronal loop implosion}
\begin{document}
\title{Simulating coronal loop implosion and compressible wave modes \\ in a flare hit active region}

\author{Aveek Sarkar${^\dag}$}
\email[$^\dag$]{aveeks@prl.res.in}
\affil{Astronomy and Astrophysics Division, Physical Research Laboratory, Ahmedabad 380009, India}

\author{Bhargav Vaidya$^\ddag$}
\email[$^\ddag$]{bvaidya@iiti.ac.in}
\affil{Centre of Astronomy, Indian Institute of Technology Indore, Simrol, Indore 453552, India}

\author{Soumitra Hazra$^\S$}
\email[$^\S$]{soumitra.hazra@gmail.com}
\affil{Center of Excellence in Space Sciences India, Indian Institute of Science Education and Research Kolkata, Mohanpur 741246, India \\
Department of Physics and Astronomy, Georgia State University, Atlanta, GA 30303, USA}

\author{Jishnu Bhattacharyya$^\P$}
\email[$^\P$]{jishnub@gmail.com}
\begin{abstract}
There is considerable observational evidence of implosion of magnetic loop systems inside solar coronal active regions following high energy events like solar flares. In this work, we propose that such collapse can be modeled in three dimensions quite accurately within the framework of ideal magnetohydrodynamics. We furthermore argue that the dynamics of loop implosion is only sensitive to the transmitted disturbance of one or more of the system variables,~\emph{e.g.} velocity generated at the event site. This indicates that to understand loop implosion, it is sensible to leave the event site out of the simulated active region. Towards our goal, a velocity pulse is introduced to model the transmitted disturbance generated at the event site. Magnetic field lines inside our simulated active region are traced in real time, and it is demonstrated that the subsequent dynamics of the simulated loops closely resemble observed imploding loops. Our work highlights the role of plasma $\beta$ in regards to the rigidity of the loop systems and how that might affect the imploding loops' dynamics. Compressible magnetohydrodynamic modes such as kink and sausage are also shown to be generated during such processes, in accordance with observations.
\end{abstract}
\section{Introduction}
Loop systems in the solar coronal active regions go through various dynamical processes following solar transient events like flares, jets, etc. By now, we have accumulated ample evidence of such dynamical processes via high resolution spaceborne observations \citep{markus99,nakariakov99, wang04, abhishek08, joshi09, gosain12,simoes13, dipu16}. A particularly interesting process among these, called~\emph{coronal loop implosion}\footnote{In this paper, the term `loop implosion' primarily signifies the observed kinematical collapse of solar coronal loop-systems and~\emph{not} the initial disturbance (\emph{e.g.}, solar flares or jets) that may initiate such collapse.}, is when the loop systems above the epicenters collapse downwards \citep{gosain12,simoes13}. In fact, observations indicate that loop systems undergo a sequence of expansions and contractions following the initial implosion. Additionally, such flare hit loop systems exhibit the propagation of conventional magnetohydrodynamic (MHD) compressible modes such as kink \citep{markus99,nakariakov99} and/or sausage \citep{abhishek08,abhidurg13}.

Despite the abundance of observational evidence, a proper understanding of the physics of such coronal loop implosions is still lacking. In his seminal work,~\citep{hudson00} conjectured that the reduction of volume integrated coronal magnetic energy between the initial and final static states could be responsible for such implosions. It was further anticipated that the most likely location where such magnetic energy reduction may take place are magnetic reconnection sites where the flare is originally sourced. This conjecture was later invoked to explain various observations~\citep{russel15,wang16}. More recently,~\cite{zuccarello17} have shown using a three dimensional zero $\beta$ MHD model that vortices developed close to the footpoints of the loop during a flare eruption may cause the loop to implode. \cite{dudik17} have shown that such implosion can also be seen at the active region boundaries.

In this paper, we would like to propose that it is actually advantageous to separate the study of coronal loop implosion (\emph{i.e.}, `the effect') from various high energy phenomena, \emph{e.g.} solar flares, jets etc. that may initiate such loop implosions (\emph{i.e.}, `the causes'). We want to argue that~\emph{the dynamics of loop implosion is only sensitive to the transmitted disturbance generated at the event site}. Towards that end, we exclude the site of the high energy activity (perhaps located at the solar photosphere) from our simulated portion of the active region. Rather in this work, we model the effect of one such disturbance by a velocity pulse generated at the event site that eventually hits the studied loop-system during the initial phase of the simulation. We believe that this viewpoint is particularly helpful while focusing on those aspects of loop dynamics which are most relevant observationally.

Within this setup, the goal of the present paper is to demonstrate that~\emph{coronal loop implosion is primarily an ideal MHD phenomenon}. Indeed, through our simulations we were capable of demonstrating implosion of coronal loops and subsequent generation of various MHD modes, in reasonable quantitative agreement with observations. Consequently, `non-ideal' dissipative MHD effects such as magnetic reconnection may, at best, play a secondary role in shaping these aspects of loop implosion dynamics.

The rest of this paper is organized as follows: in section~\ref{sec:setup} we describe the numerical setup of the problem. Our simulation results are analyzed as well as compared and contrasted with spacecraft observations in section~\ref{sec:analysis}. In particular, subsection~\ref{sec:analysis:collapse} is devoted to the study of collapse of simulated loops following a flare-like event, while subsection~\ref{sec:analysis:modes} describes compressible magnetohydrodynamic modes like kink and sausage that are generated thereafter. We summarize and discuss our setup and results further in section~\ref{sec:summary}.
\section{Numerical setup}\label{sec:setup}
	\begin{figure}
	\includegraphics[width=0.48 \textwidth,angle=0]{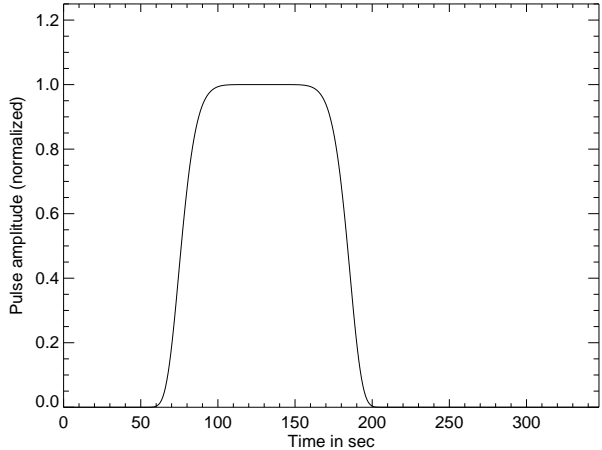}
	\caption{Time profile of the velocity pulse applied at the bottom boundary. \\}
	\label{pulse}
	\end{figure}
In this work, we have modeled an active region of the solar corona inside a three dimensional gravitationally stratified rectangular box and have solved the compressible MHD equations for an adiabatic environment~\citep{nakariakov05} inside the box using the PLUTO code \citep{mignone07}. The box was considered at $70$ Mm above the solar photosphere with horizontal ($x$ and $y$ directions) and vertical ($z$ direction) extents of $350$ Mm (each) and $700$ Mm respectively. The resolution of the box was taken to be $256 \times 256 \times 512$.

For reasons to be clarified in the next section, we ran two simulations inside the above mentioned box with two different configurations for the initial magnetic field. In both cases, a potential dipole \citep{ofman02} was introduced at the base of the box (\emph{i.e.}, embedded at the photosphere, $70$~Mm beneath the box) to create part of the synthetic active region. An additional constant magnetic field, representing Sun's own dipolar magnetic field in the vicinity of the active region under consideration, was introduced on top of the above in one of the simulations, while the other simulation was initiated just with the original dipolar field. Further details about the magnetic fields can be found in the following section.

A velocity pulse along the vertical $z$-direction, emulating the effect of the transmitted disturbance due to a high energy event, was introduced from the bottom boundary (precise location: $-42$ Mm $\leq x \leq 42$ Mm, $-42$ Mm $\leq y \leq 0$ Mm, $z = 70$ Mm)~\citep{ofman02}. The velocity pulse started at $52$~s, continued until $205$~s, and maintained a time profile of the form $V_{z}(t) = V_{z}(0)\exp\left(-[(t-t_0)/\delta t]^8\right)$, where $V_{z}(0)$ is the amplitude of the pulse, $t_0 = 130.1$~s is the mid-time of the pulse and $\delta t = 56.4$~s is the half-width of the pulse. The pulse characteristics was chosen in accordance with~\cite{ofman07}, and a normalized time profile of it is shown in Figure~\ref{pulse}. Observations indicate~\citep{markus99} that plasma propagates with an outward velocity of about $700$ - $1000$~km/s at the flaring site. Complying with such observations, the prescribed velocity amplitude at the source was set to be $V_{z}(0) = 0.012 \times V_a$, where $V_a = 8070$ km/s is the velocity scale of the system.

The simulations were initiated with a static atmosphere where downward gravity force balances the upward pressure gradient force. Since the dipole is potential, and the constant magnetic field added on top of it in one of the simulations is also current free, there is no initial Lorentz force affecting the initial equilibrium configuration in either simulation. Hence, the initial pressure and density profile are identical in both the simulations; their spatial variation is shown in the appendix below.

Free flowing boundary conditions for all variables were imposed on all four boundaries along the $x$ and $y$ directions. At the top boundary, density and pressure were fixed to their initial values, whereas boundary values for all three components of velocity and magnetic fields were copied from the last domain cell in the vertical direction. At the bottom boundary, velocity components were set to zero except throughout the duration of the pulse. Density and pressure at the bottom boundary were fixed to their initial values and standard outflow boundary conditions were imposed on all components of the magnetic field.
\section{Analysis}\label{sec:analysis}
	\begin{figure}
	\includegraphics[width=0.48 \textwidth,angle=0]{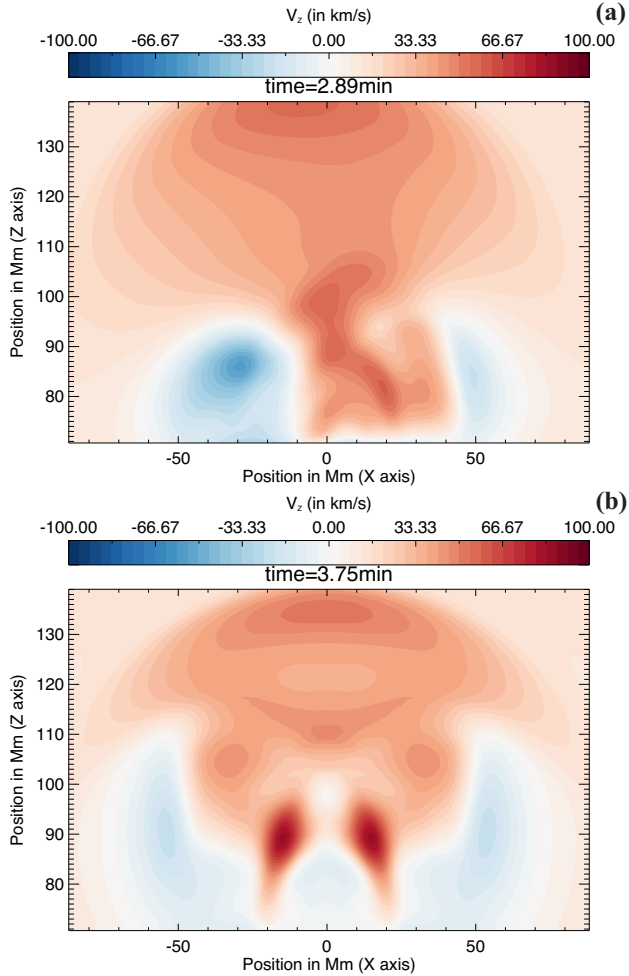}
	\caption{Snapshots of propagation of the $V_z$ disturbance through the active region on the $xz$ plane ($y = 0$) at $t = 2.89$ min (\textbf{panel (a)}:~\texttt{simulation-1}) and at $t = 3.75$ min (\textbf{panel (b)}:~\texttt{simulation-2}). Colored versions of these figures along with the corresponding movies (\href{https://youtu.be/vOuVvbrW5aw}{\texttt{fig2amovie.mp4}} and \href{https://youtu.be/j4RXqDqtncM}{~\texttt{fig2bmovie.mp4}}, respectively) are available online.\\}
	\label{velz}
	\end{figure}
As indicated previously, we ran two simulations with different configurations for the initial magnetic field. In one of these cases (henceforth \texttt{simulation-1}), the initial magnetic field was set up to include two separate components:
	\begin{enumerate}
	\item a `local' dipolar part $\mathbf{B}_\textsc{d}$, of strength $0.0117$ Tesla at the photosphere, whose field lines were `pegged' on the photosphere;
	\item a `global' component $\mathbf{B}_\textsc{g}$, which was a combination of constant fields of $0.000468$ Tesla in both horizontal ($x$ and $y$) directions.
	\end{enumerate}
$\mathbf{B}_\textsc{g}$ approximated Sun's own large scale dipolar magnetic field within the active region. It is justified to assume this to be a constant, since its spatio-temporal variation is much smaller than the size of the active region. The net magnetic field $\mathbf{B}$ of \texttt{simulation-1} was thus given as $\mathbf{B} = \mathbf{B}_\textsc{d} + \mathbf{B}_\textsc{g}$.

The other simulation (henceforth \texttt{simulation-2}) only included the local dipolar part $\mathbf{B}_\textsc{d}$ (field strength as above) such that its total magnetic field was $\mathbf{B} = \mathbf{B}_\textsc{d}$. Since solar active regions are embedded in the global dipolar magnetic field of the Sun, the former configuration emulated the active region under consideration more accurately than the latter case.

The inclusion of Sun's global dipolar field in \texttt{simulation-1} contributed by enhancing the local magnetic energy density and overall decreased the value of local plasma $\beta = 8\pi p/B^2$. This change in profile of $\beta$ could be understood a bit more quantitatively in the initial configurations, where the profiles of every field are known analytically. In both simulations, given the stratified atmosphere, the pressure roughly falls off as fast as the gravitation potential (since the gradient of one balances that of the other). On the other hand, the synthetic stand-alone dipole magnetic field $\mathbf{B}_\textsc{d}$ is ought to fall off as $r^{-3}$, $r$ being the distance of the plasma from the location of the dipole. Therefore, in the absence of the constant global solar dipolar field $\mathbf{B}_\textsc{g}$, plasma $\beta$ must increase much more rapidly with height than in reality. Inclusion of the global scale dipolar field of the Sun thus contributes significantly to bring down the active region plasma $\beta$.

Naturally, one may be curious about the physical significance and/or relevance of \texttt{simulation-2}, where plasma $\beta$ goes up to perhaps unrealistically large values. The answer lies in our primary goal to separately focus on the dynamics of loop implosion by disentangling it from the multiple causes that may initiate it. Studied as an ideal MHD phenomena, it is worthwhile to understand the correlation between the character of a given collapse with the local plasma $\beta$ of the medium. This has prompted us to allow plasma $\beta$ to vary even outside its `realistic' range. Hence, many of the following results come from \texttt{simulation-2} where $\beta$ is allowed to be quite large. In this regard, the following two points are worth emphasizing:
	\begin{enumerate}
	\item \emph{qualitative} aspects of loop collapse are hardly affected by higher values of $\beta$;
	\item as long as we focus on parts of the active region where $\beta$ stays within its acceptable range, physical quantities like collapse rates computed in \texttt{simulation-2} are in rather remarkable agreement with observations, even though the local $\beta$ could be unrealistically high in some other parts of the active region here.
	\end{enumerate}
Both these statements will find support from explicit numerical results presented below.

We ran \texttt{simulation-1} for about $15$ mins, while \texttt{simulation-2} was run for about $35$ mins (for reasons to be addressed in the final discussions; see section~\ref{sec:summary}). As mentioned before, the initial velocity pulse effectively encoded the disturbance originated at the solar photosphere, and affected the embedded magnetic field of the box. The pulse developed a magnetosonic wave in the ambient magnetic field and propagated through the medium, waxing and waning it. The reader may seek further clarification from the online movies \href{https://youtu.be/vOuVvbrW5aw}{(\texttt{fig2amovie.mp4}} for \texttt{simulation-1} and \href{https://youtu.be/j4RXqDqtncM}{\texttt{fig2bmovie.mp4}} for \texttt{simulation-2}, respectively) on the time evolution of the $V_z$ on the $xz$-plane (at $y = 0$); snapshots from the movies are shown in Figure~\ref{velz}.

In the following sections, we will first address the dynamical response of the simulated loops to the disturbance induced by the velocity pulse. Subsequently, the compressible MHD modes that gets generated will be described.
\subsection{Loop collapse}\label{sec:analysis:collapse}
	\begin{figure}
	\includegraphics[width=0.48 \textwidth,angle=0]{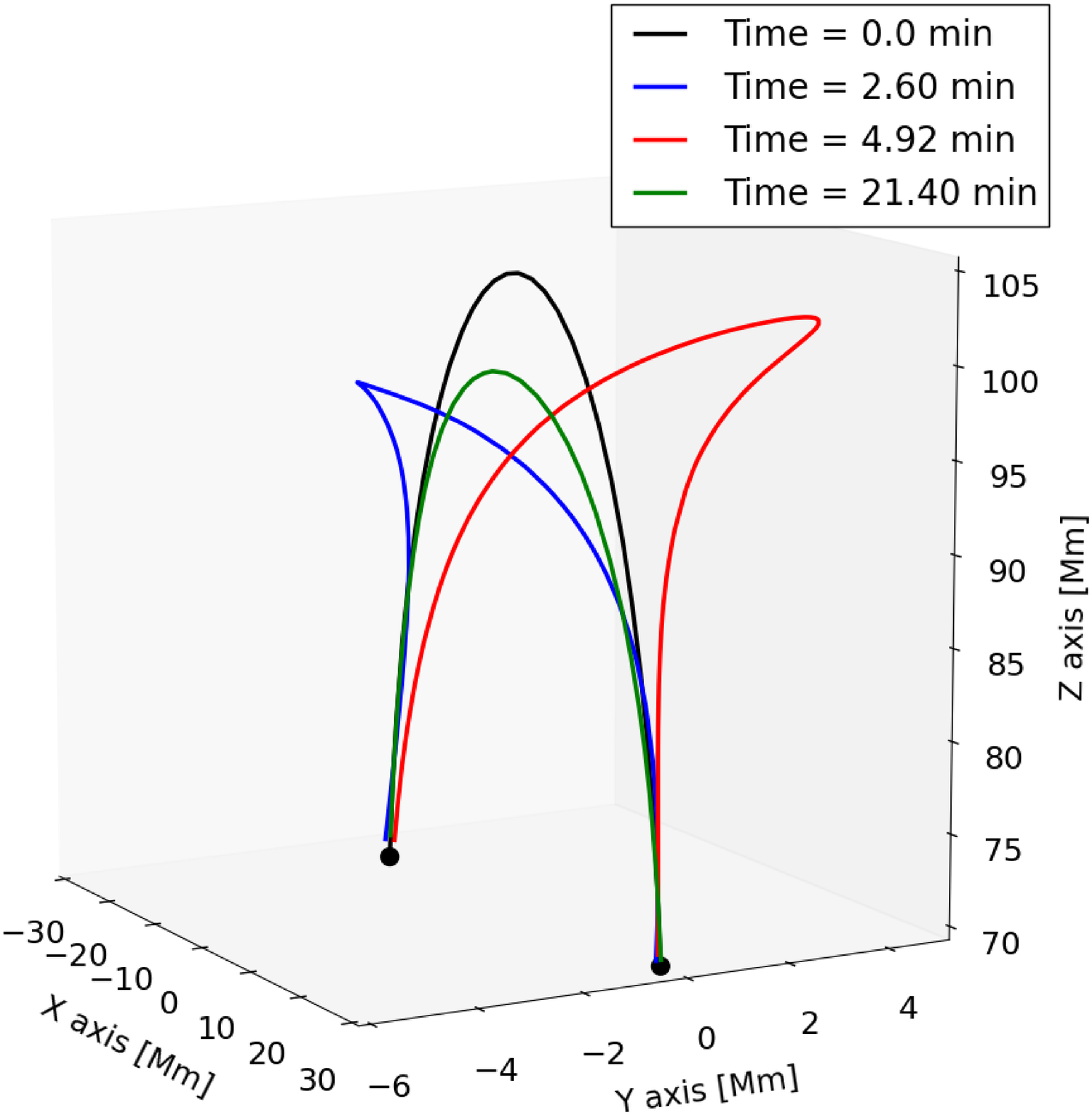}
	\caption{Orientation of a typical collapsing loop from~\texttt{simulation-2} seeded at $\{25.9$~Mm, $0$~Mm, $70$~Mm$\}$ at different moments (exact times in inset). A colored version of this figure along with a movie are available \href{https://youtu.be/z1wobrLma-M}{online}.\\}
	\label{kink}
	\end{figure}

	\begin{figure}
	\includegraphics[width=0.48 \textwidth,angle=0]{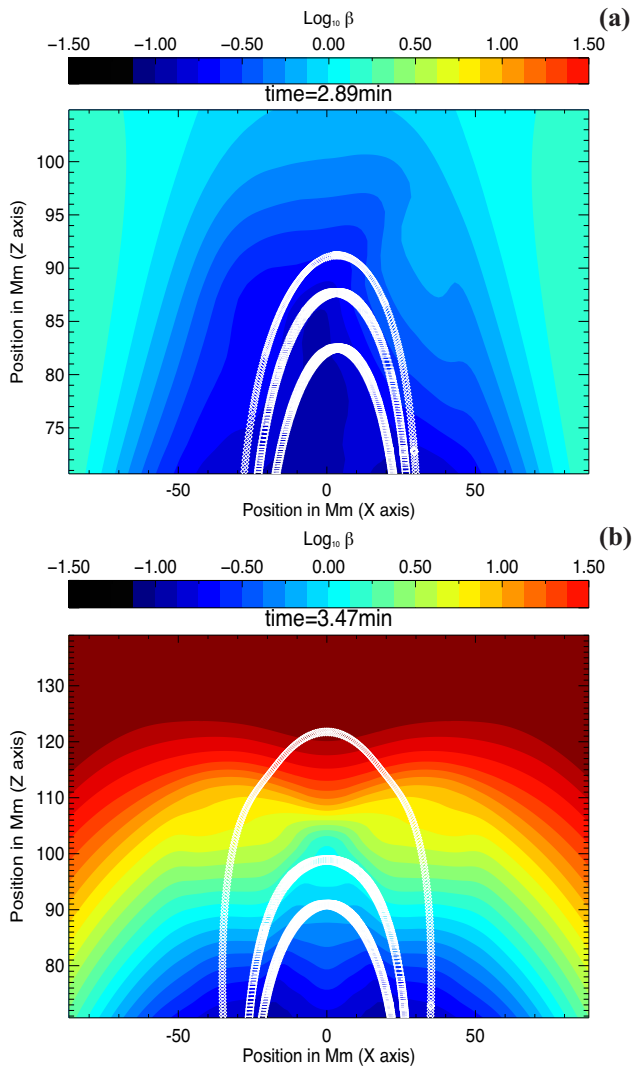}
	\caption{\textbf{Panel (a):} Snapshot of variation of plasma $\beta$ on the $xz$ plane (at $y = 0$) at $t = 2.89$ min from~\texttt{simulation-1} (taken from movie: \href{https://youtu.be/xiPncEXwTtU}{\texttt{fig4amovie.mp4}}). The white curves denote projections of the instantaneous configurations of three loops seeded at $\{21$~Mm, $0$~Mm, $70$~Mm$\}$, $\{25.9$~Mm, $0$~Mm, $70$~Mm$\}$ and $\{29.4$~Mm, $0$~Mm, $70$~Mm$\}$ respectively. \textbf{Panel (b):} Snapshot of variation of plasma $\beta$ on the $xz$ plane (at $y = 0$) at $t = 3.47$ min from~\texttt{simulation-2} (taken from movie: \href{https://youtu.be/Rpi7gjBn0zM}{\texttt{fig4bmovie.mp4}}). The white curves denote projections of the instantaneous configurations of three loops seeded at $\{22$~Mm, $0$~Mm, $70$~Mm$\}$, $\{25.9$~Mm, $0$~Mm, $70$~Mm$\}$ and $\{35$~Mm, $0$~Mm, $70$~Mm$\}$ respectively. The middle curve from this figure is a projection of the loop shown in Figure~\ref{kink}. \\ \-\ \-\ Colored versions of the above figures along with the corresponding movies are available online.\\}
	\label{plasmabeta}
	\end{figure}

To explore the effect of the pulse on the magnetic field lines, in both the simulations, we selected multiple points at the base of the box ($z = 70$ Mm) in and around the velocity pulse site. Taking these points to represent seed footpoints of the loops, we traced the corresponding magnetic field lines through them until we reached the other footpoints. These field lines thus represent the simulated coronal loops/strands. Evolution of these field lines were then tracked at regular time intervals. As a typical example, Figure~\ref{kink} plots the configuration of one such loop from~\texttt{simulation-2} at three instances of time. For actual depiction of the evolution of the same loop in time, we refer the reader to the corresponding online movie (\href{https://youtu.be/z1wobrLma-M}{\texttt{fig3movie.mp4}}).

As stated before, exploring the correlation between the characteristics of loop implosion with the local plasma $\beta$ profile of the system was among our goals in this project. Towards that end, the online movies\href{https://youtu.be/Rpi7gjBn0zM}{~\texttt{fig4amovie.mp4}} and~\href{https://youtu.be/Rpi7gjBn0zM}{\texttt{fig4bmovie.mp4}} capture the time evolution of plasma $\beta$ of the system in~\texttt{simulation-1} and~\texttt{simulation-2}, respectively. These movies also depict the time evolution of the projection of some of the traced loops in these simulations; Figures~\ref{plasmabeta}(a) and~\ref{plasmabeta}(b) are snapshots from these movies, respectively.

\begin{figure}
	\includegraphics[width=0.48 \textwidth,angle=0]{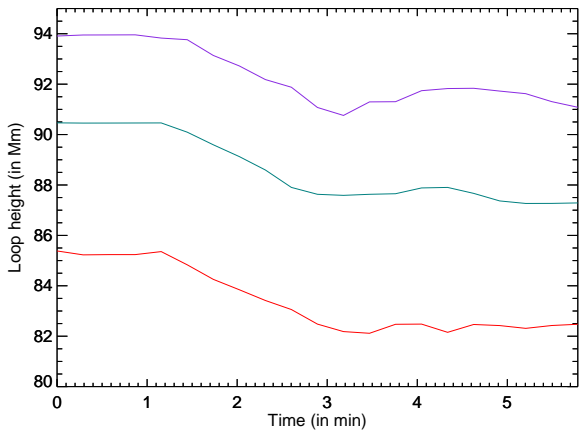}
	\caption{Time evolution of the height of the midpoint (\emph{i.e.}, the $z$ coordinate of the midpoint) of three loops from~\texttt{simulation-1} whose projections appear in panel (a) of Figure~\ref{plasmabeta}.\\}
	\label{implosion:sim-1}
	\end{figure}

As seen from these movies and plots, in both simulations, the loops inside the box start to contract soon after the release of the velocity pulse, followed by a brief expansion phase and an eventual further steady shrinkage. Similar features have been pointed out in solar coronal loops observed during flares using the Solar Dynamic Observatory \citep{sun12,gosain12}. In fact, it is customary to trace the looptops during spacecraft image analysis. To facilitate a quantitative comparison with such observations, we also monitor the time evolution of the midpoints of the loops, corresponding to~\texttt{simulation-1}, whose projections appear in Figure~\ref{plasmabeta}(a). Figure~\ref{implosion:sim-1} graphs this time variation, clearly indicating implosion of these loops in a manner consistent with observations \citep{liu12, gosain12, simoes13, shen14, russel15, wang16}.

\begin{figure}
	\includegraphics[width=0.5 \textwidth,angle=0]{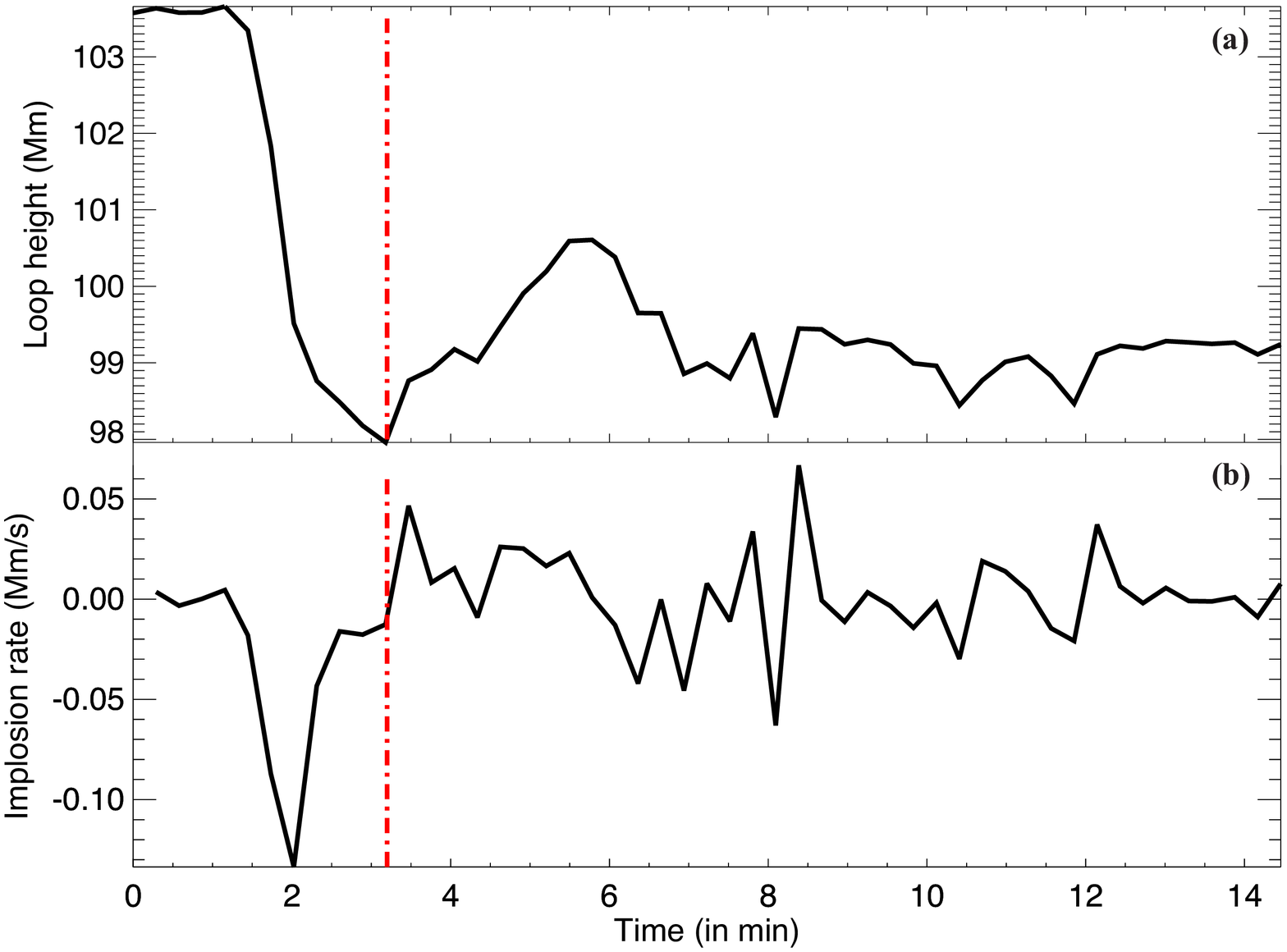}
	\caption{\textbf{Panel (a):} Time evolution of the height of the midpoint (\emph{i.e.}, the $z$ coordinate of the midpoint) of the loop from~\texttt{simulation-2} shown in Figure~\ref{kink}. The loop undergoes steady implosion until about $3.2$ min, marked by the red dotted line, after which it expands for a while and then implodes again, but both at a relatively slower rate. \textbf{Panel (b):} Variation of the implosion rate of the same loop over time.\\}
	\label{implosion:sim-2}
	\end{figure}

Similarly, we also track the motion of the midpoint of the loop shown in Figure~\ref{kink} (corresponding to~\texttt{simulation-2}) whose projection appear as the middle curve in Figure~\ref{plasmabeta}(b), and additionally evaluate its implosion/collapse rate over time. Note that plasma $\beta \lesssim 1$ along this loop, even though $\beta$ can be `rather large' in some other parts of the box in this simulation. Nevertheless, the plots of the implosion rate for this loop, shown in Figure~\ref{implosion:sim-2}, posses rather impressive similarity with the spacecraft observation of~\cite{simoes13};~\emph{e.g.}, compare F\textsc{ig}. 4 of~\cite{simoes13} with the part of Figure~\ref{implosion:sim-2} to the left of the red dotted line. This last observation strongly indicates that the implosion rate of a loop is most (if not only) sensitive only to the~\emph{local} plasma $\beta$. For example, in Figure~\ref{implosion:sim-2} the implosion rate is found to reach up to $130$ km/s, although this rate may depend on the loop under consideration (more precisely, on the local plasma $\beta$ along the loop).

A qualitative understanding of the physics underlying loop implosions could be achieved by focusing on the variation of plasma $\beta$ in space and time (see Figures~\ref{plasmabeta}(a)and (b) and the associated online movies~\href{https://youtu.be/xiPncEXwTtU}{\texttt{fig4amovie.mp4}} and~\href{https://youtu.be/Rpi7gjBn0zM}{\texttt{fig4bmovie.mp4}}). Consider, for example, the loop in Figure~\ref{kink} for concreteness (recall that the implosion rate was seen to agree fairly well with observations for this loop). For this loop, $\beta$ varies over a realistic range from $\beta \sim 0.01$ at the base of the box to $\beta \sim 1$ at the looptop according to Figure~\ref{plasmabeta}(b). Consequently, hydrodynamic forces are comparable to the local Lorentz force towards the looptop, while the latter substantially dominates over the former towards the footpoints. This also makes the loop significantly more rigid closer to the footpoint than the looptop. As the magnetosonic wave induced by the velocity pulse hits the loop, the response at the looptop is considerably more prominent (especially due to the hydrodynamic forces) than near the footpoint, deforming the loop both horizontally and vertically and making the looptop implode.

The effect is rather `exaggerated' for larger loops,~\emph{e.g.}, the largest loop in Figure~\ref{plasmabeta}(b). While $\beta$ at the looptop for such a loop is perhaps higher than what is anticipated, it illustrates the point being made here more effectively. Due to much higher $\beta$ towards the looptop, simple hydrodynamics provides a good effective description of the fluid there, and the loop is significantly slacker high up. The hydrodynamic force wave induced by the velocity pulse predominantly affects the fluid near the looptop, as can be verified explicitly. Due to magnetic flux being frozen inside the fluid elements within the realm of ideal MHD, the local magnetic field lines start to collapse as soon as the local fluid elements get pushed around by the hydrodynamic forces. One may presumably think of this as an~\emph{inverse buoyancy effect}~\citep{parker55}. Admittedly, the implosion rate for a loop like this may not be observationally relevant. However, this example is perhaps better tailored to illustrate the underlying physics more effectively. It also helps to establish our earlier claim that~\emph{qualitative} aspects of loop implosion are not too sensitive even if $\beta$ is `too large'.

On the other end, much smaller loops are more uniformly taut and rigid throughout, since change of $\beta$ from the footpoint to the looptop is comparatively less; \emph{e.g.}, for the loops in Figure~\ref{plasmabeta}(a) or the smallest loop in Figure~\ref{plasmabeta}(b) $\beta \sim 0.4$ or less at the looptop. Consequently, their post-flare deformation is significantly less pronounced;~\emph{e.g.}, see Figure~\ref{implosion:sim-1}. Upon comparing the implosion rates of all the loops in Figure~\ref{plasmabeta} (both panels), one also finds the following trend: looptops of bigger loops achieving higher $\beta$ have higher implosion rates. This trend indicates that successively smaller (and observationally significant) loops would eventually cease to implode but rather effectively oscillate vertically. This should be identified with the~\emph{vertical kink oscillations} observed in some of the post-flare loop systems \citep{wang04, kim14}. Extrapolating the trend even further, it is natural to anticipate a cutoff on the size of the loops below which the loops would not show any observationally significant post-flare deformation. Almost similar trend is observed in~\cite{simoes13}.

The movies also indicate that a looptop, initially submerged in a higher $\beta$ environment, never regains its original height after imploding. Rather, it oscillates for a while about its lower final height before settling down in a lower $\beta$ surrounding. This behavior has precisely been identified using Solar Dynamic Observatory observations \citep{liu12, simoes13}, and an explanation of the phenomena was attempeted by \cite{russel15}.
\newpage
\subsection{Sausage and kink modes}\label{sec:analysis:modes}
	\begin{figure}
	\includegraphics[width=0.5 \textwidth,angle=0]{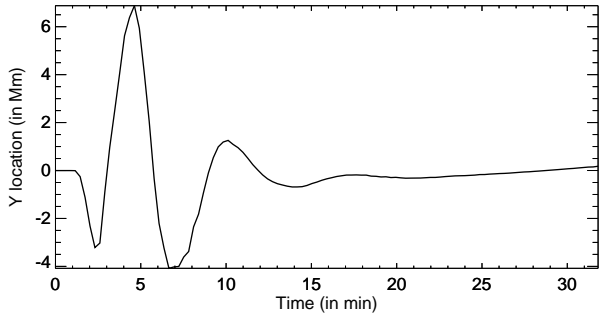}
	\caption{Variation of the $y$ coordinate of the mid-point of the collapsing loop from~\texttt{simulation-2} shown in Figure~\ref{kink} demonstrates the existence of the fundamental kink mode. The plot shows that the loop oscillates for almost 2 periods.\\}
	\label{kink1d}
	\end{figure}
The simulated imploding loop systems also show the presence of the fundamental kink mode $(k_z = 0$, $m = 1)$~\citep{roberts83}. We saw the orientation of a typical traced loop seeded near the epicenter at four different instances in Figure~\ref{kink}. The transverse deformation of the loop clearly depicts the existence of the fundamental kink mode. The $y$ coordinate of the midpoint of the loop is plotted over time in Figure~\ref{kink1d}. This is seen to oscillate in time for almost two periods with a periodicity of about $5$ - $7$ min. Considering the average length of the loop to be $89.5$ Mm (between $3.18$ min and $8.38$ min), the speed of the mode is estimated to be $\sim 600$ km/s, which is less than the loop integrated average Alfv\`en speed ($\sim 800$ km/s) but more than the average sound speed ($330$ km/s) along the loop over the same period of time. All other traced loops seeded in the vicinity of the epicenter also demonstrate similar kink mode oscillations.

	\begin{figure}
	\includegraphics[width=0.5 \textwidth,angle=0]{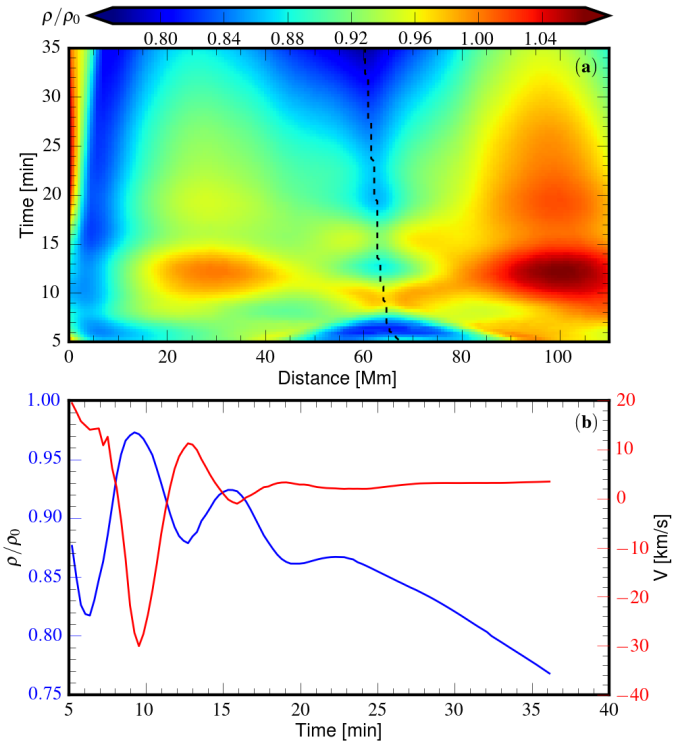}
	\caption{\textbf{Panel (a):} Density evolution along a strand from~\texttt{simulation-2} seeded at $\{35$~Mm, $-21$~Mm, $70$~Mm$\}$ shows density variation indicative of the standing sausage wave oscillation. Since the loop shrinks over time, range on the horizontal axis is limited by the length of the shortest loop. The dotted line depicts the location of the looptop. \textbf{Panel (b):} Evolution of looptop density and component of velocity along the loop over time. Density shows anti-correlation with the velocity which is a typical characteristic of the sausage wave. A colored version of this figure is available online.\\}
	\label{sausage}
	\end{figure}
The time evolution of the density along another typical loop from~\texttt{simulation-2}, also seeded at the velocity injection site, is shown in Figure~\ref{sausage}(a). One may notice that different parts of the loop gets overdense and rarefied periodically. This is a clear signature of the fundamental sausage mode oscillation, another axisymmetric magnetosonic mode $(k_z = 0$, $m = 0)$ which makes the outer loop surface (and thus the area of cross section) oscillate in a periodic manner. Figure~\ref{sausage}(b) is a time-density plot through the mid point of the loop. It shows that the periodicity with which the looptop becomes overdense is about $6.5$ - $7$ min. The typical length of the selected loop was calculated to be $\sim 127$ Mm, which makes the phase speed to be around $600$ - $650$ km/s. This is greater than the loop integrated average Alfv\'en ($445$ km/s) and the sound ($324$ km/s) speeds during the first oscillation cycle. To further confirm the presence of these modes in our simulated loop system, the component of the local velocity field along the loop is over-plotted in Figure~\ref{sausage}(b); this demonstrates that the velocity component oscillates $90^{\circ}$ out of phase with the density~\citep{tsuneta09}.
\section{Summary and discussions}\label{sec:summary}
In this work, we have presented one of the first attempts ever to study the implosion of solar coronal loops by tracing magnetic field lines in a simulated flare hit solar active region. By tracking these field lines over time, we were able to demonstrate various observed features of a post-flare loop system. The central theme of this work could be summarized through the following two points:

\paragraph{Physics of implosion of coronal loops should be studied separately from its cause:} We have proposed that the dynamics of imploding loop systems, especially the observationally relevant aspects of it, could be most efficiently explored by focusing on the loop system itself. The present work seems to indicate that loop collapse is rather insensitive to the precise nature of the high energy phenomenon occurring at the solar photosphere (\emph{e.g.}, jets, sprays etc. beside solar flares); rather it would get initiated by,~\emph{e.g.}, a velocity disturbance that results from such an event. Therefore, instead of including the explosion site into the simulation, it is more sensible to model the effect of the explosion through a disturbance,~\emph{e.g.}, in the velocity profile. Of course, the amplitude and the direction of the velocity pulse would encode the nature and intensity of the initial event occurring at the photosphere, and would presumably affect the implosion rate among other things. Also, such an effect could be equivalently modeled through an initial disturbance of some other field variable,~\emph{e.g.}, pressure or magnetic field, instead of introducing an initial velocity pulse. We leave such studies for the future.

\paragraph{Coronal loop implosion is predominantly an ideal MHD phenomenon:} We have further proposed that ideal MHD furnishes an adequate framework to  model the dynamics of imploding solar coronal loops, without the need to consider the region of free energy release as part of the simulated region (in accordance with the previous point). Within this setup, we have tried to establish that local implosion characteristics (\emph{e.g.} implosion rates), are primarily governed by local plasma characteristics (\emph{e.g.} local plasma $\beta$ profile).

In support of the above claims, we have modeled a portion of the active region $70$~Mm above the solar photosphere. We have subsequently demonstrated that as long as the immediate neighborhood of the loops admit `realistic' values of $\beta$, simulated loop implosion is in remarkable agreement with observations~\citep{liu12, gosain12, simoes13, wang16}. Overall, the present simulations show that the value of $\beta$ in a region affects the rigidity of the local loop system substantially. This determines their lateral and vertical movement in turn, giving rise to a host of interesting phenomena including loop implosion as well as horizontal and vertical kink oscillations.

Extrapolating these observations further, one would thus naturally expect that loop implosions should be fairly generic phenomena, and should be observed when significant disturbances originating from any kind of source propagate through parts of the solar corona admitting relatively high plasma $\beta$. Of course, every flare hit active region may not demonstrate the existence of imploding loops, perhaps because identifying loops with relatively high plasma $\beta$ is not always feasible as they are relatively higher up in the corona. However, having observed such implosion, it is conceivably possible to estimate the values of the local plasma $\beta$ from the observed implosion rate.

\paragraph{Imploding loops also show the signature of compressible MHD waves:} Loops seeded in the vicinity of the epicenter also demonstrate the propagation of both sausage and kink modes spontaneously, as expected. Even though we primarily focused on describing the fundamental kink mode in this article, higher harmonics in simulated loops are visible when loops enter `relatively higher' $\beta$ regions and become more supple. While this connection between $\beta$ and the harmonics number could be observationally significant, identifying higher harmonics observationally may be difficult \citep{brady07} even with present generation instrumentation.

Admittedly, the background of~\texttt{simulation-2} supports a plasma $\beta$ profile with a rather steep gradient than what is observed. However, as we have stated earlier (1) most of the~\emph{qualitative} aspects of loop implosion are not too sensitive to this issue, and (2) inclusion of the global solar dipolar field rectifies this problem by bringing down $\beta$ substantially. Of course, to be observationally relevant a simulation of coronal loop implosion should be performed with higher resolution. In this light, the present work should be considered more as a `proof of concept', allowing us to establish the claims made earlier in this section, within a very simple computational setting and without giving up any essential physics. The role of \texttt{simulation-1} was to barely convince ourselves that the higher values of $\beta$ in \texttt{simulation-2}, even though `unrealistic', did not affect our main conclusions. This is also why \texttt{simulation-2} was run roughly twice as long as \texttt{simulation-1} as noted earlier, and the various MHD oscillations were also studied using \texttt{simulation-2}.

Another key property of major concern is damping of loop oscillations~\citep{ruderman02,pascoe16}. In general, Braginskii viscosity~\citep{braginskii65} of the coronal plasma is incapable of explaining the observed damping rate of loop oscillations. We suspect that Lorentz force may act as a damping agent due to the relatively low value of $\beta$ near the footpoints. However, further numerical simulations are required to establish this claim.
\appendix
\section{Initial conditions: pressure and density profiles}
In both the simulations reported in this work, the initial state is a static (\emph{i.e.}, all time derivatives vanish, so does the velocity field) solution of the MHD equations by assumption where the magnetic field is current free by choice. Hence, the density and pressure profiles -- which are themselves related via the adiabatic equation of state $p = \rho^\gamma$ in normalized unit -- are determined via the static momentum balance equation and are identical in both the simulations. Translation invariance along the horizontal $x$ and $y$ directions make the pressure and density constant along these directions. On the other hand, since gravity acts vertically along the $z$ direction, the static non-dimensional momentum balance equation along this direction is given by
	\begin{equation}\label{static}
	-\frac{\xi}{2}\frac{\mathrm{d}p}{\mathrm{d}z} - \frac{\rho}{F_r(R_0 + z - z_\text{min})^2} = 0~,
	\end{equation}
where $R_0$ is the radius of the Sun, $z$ is the vertical distance of a fluid element from the solar photosphere in non-dimensional units, $z_\text{min}$ is the height of the bottom boundary of the box from the solar surface, $F_r$ is called the Froude number defined as $F_r = V_{a}^2 R_0/G M_{\odot}$, where $G$ is universal gravitational constant, $V_a = 8070$~km/s is the system's Alfv\'en speed and $M_{\odot}$ is the mass of the Sun, and finally $\xi = (2/\gamma)(c_s/V_{a})^2$, $c_s$ being the characteristic sound speed of the system. Solving equation~\eqref{static}, we get the gravitationally stratified hydrostatic density and pressure profiles as
	\begin{equation}
	\begin{split}
	\rho & = \rho_0\left[1 + \frac{1}{H}\frac{\gamma - 1}{\gamma}\frac{\rho_0}{p_0}\left(\frac{1}{R_0 + z - z_\text{min}} - \frac{1}{R_0}\right)\right]^{\displaystyle{\frac{1}{\gamma - 1}}}~, \\
	& \\
	      p & = p_0\left[1 + \frac{1}{H}\frac{\gamma - 1}{\gamma}\frac{\rho_0}{p_0}\left(\frac{1}{R_0 + z - z_\text{min}} - \frac{1}{R_0}\right)\right]^{\displaystyle{\frac{\gamma}{\gamma - 1}}}~,
	\end{split}
	\end{equation}
where, $p_0 = 108.0$~Pa, $\rho_0 = 1.673 \times 10^{-18}$~kg/m$^3$ and $H = 2/\xi F_r$ is the normalized scale height.
\begin{acknowledgments}
Computations were carried out on the Physical Research Laboratory's VIKRAM cluster. We are grateful to an anonymous referee for providing us with very useful feedback. S.H. likes to acknowledge hospitality and financial support from PRL during his visit.
\end{acknowledgments}

\end{document}